\documentclass[showpacs,preprintnumbers,superscriptaddress]{revtex4}
\usepackage[T1]{fontenc}
\usepackage{amsfonts}
\usepackage{amsmath,amsbsy,amssymb,graphicx} 
\usepackage{times}
\usepackage{color}

\begin{document}                
     
\title{Dicke Model for Quantum Hall Systems}  
\author{Y.~Hama}
\affiliation{RIKEN Center for Emergent Matter Science (CEMS), Wako, Saitama 351-0198, Japan}
\author{M.~H.~Fauzi}
\affiliation{Department of Physics, Tohoku University, Sendai 980-8578, Japan}
\author{K.~Nemoto}
\affiliation{National Institute of Informatics, 2-1-2 Hitotsubashi, Chiyoda-ku, Tokyo 101-8430, Japan}
\author{Y.~Hirayama}
\affiliation{Department of Physics, Tohoku University, Sendai 980-8578, Japan}
\affiliation{WPI-Advanced Institute for Materials Research, Tohoku University, Sendai 980-8577, Japan}
\author{Z.~F.~Ezawa}
\affiliation{Advanced Meson Science Laborary, Nishina Center, RIKEN, Wako 351-0198, Japan}
\date{\today}

\begin{abstract}{ 
Quantum Hall (QH) systems consist of many-body electron  and nuclear spins.
They are coupled so weakly through the hyperfine interaction
that  normally  electron spin dynamics are scarcely affected by the nuclear spins.
The dynamics of the QH systems, however, may drastically change  when the nuclear spins interact with  
low-energy collective excitation modes of the electron spins.  
We  theoretically investigate the nuclear-electron spin interaction in the QH systems as  hybrid quantum systems driven by the hyperfine interaction. 
In particular, we study the interaction between the nuclear spins and the   Nambu-Goldstone (NG) mode
with the linear dispersion relation associated with the U(1) spin rotational symmetry breaking. 
We show that such an interaction 
is described as   nuclear spins collectively coupled to the NG mode, and
can be effectively described by the  Dicke model. 
Based on the model we suggest that various collective spin phenomena realized in quantum optical systems  also emerge in the QH systems.}
  
\end{abstract}

\maketitle

\section{Introduction}\label{sec1:introduction}
Quantum Hall (QH) systems exhibit fascinating macroscopic quantum phenomena \cite{S. Das Sarma1,Ezawa:2013ae}.
For instance, various low-energy electron coherent phenomena in terms 
of  spin and/or pseudospin (layer) internal degrees of freedom are realized. 
Research interests for the QH physics are not  limited to the electron spin physics.
However, the electron-nuclear spin dynamics has not yet attracted much attention.
For instance,  the  GaAs semiconductor with the $s$-type conduction band has a large natural abundance of nuclear spins with the  3/2 nuclear spin angular momentum ($^{69}$Ga, $^{71}$Ga, and $^{75}$As).
Although the nuclear spins interact with the electron spins mainly through the Fermi contact hyperfine interaction, 
 it is usually so weak that electron transport properties are not affected  by this interaction.
Thus, previous studies of QH physics have mainly been focused on the electron spin physics, 
whereas the nuclear spins are utilized merely as a tool to investigate the electron magnetic properties \cite{Hirayama01}. 

The above situation may  change when the nuclear spins interact with low-energy excitation modes of electron spins such as Nambu-Goldstone (NG) modes.
These modes appear in the canted antiferromagnetic (CAF) phase \cite{S. Das Sarmaetal,Pellegrinietal,Schliemannetal,Lopatnikovaetal,Ezawa:2005xi,Fukudaetal,Kumadaetal,yhamaetal,Fauziprb}, 
which is the most interesting one among the three phases of the total filling factor $\nu=2$ bilayer QH systems; 
the ferromagnetic phase, the spin-singlet phase and the CAF phase. 
In the CAF phase, the antiferromagnetic correlations between the electron spins in one of the layers (abridged
as front layer) and those in the other layer (abridged as back layer) are generated, and the associated linear dispersing NG mode emerges. 
A new physics is expected to emerge in this phase. 
In the related context, the nuclear spin relaxation was  experimentally estimated by using the 
resistivity-detected nuclear spin relaxation measurement \cite{Kumadaetal},
where the longitudinal resistance $R_{xx}$ is used as a measure of the nuclear-spin polarization.
 It has been shown that the nuclear-spin relaxation time in the CAF phase is the shortest compared with those in the other two phases  (the ferromagnetic phase and the spin-singlet phase).
More recent experiment \cite{Fauziprb},
where the spatial nuclear-spin polarization distribution was recorded after the exposure to the CAF phase, 
showed a sudden change in the nuclear-spin polarization distribution from the initial one.
These experimental results suggest a unique many-body interaction between electron and nuclear spin systems.

It is worthwhile to investigate the nuclear spin physics mediated by the hyperfine interaction in the QH systems also from the following three perspectives: 
First,  nonequilibrium phenomena of nuclear spins in the QH systems are still less understood to date.
Second, we expect a rich variety of many body effects as well as cooperative phenomena in terms of a nuclear-spin ensemble  driven by the magnetic properties of the QH systems,
just like the superradiance phenomena in  optical cavity quantum electrodynamics systems 
composed of atomic Bose-Einstein condensate \cite{cqed1} and in quantum dots \cite{qd1,qd3}.
Third,  the electron-nuclear spin hybrid system is a candidate for a spin-base quantum information processing and computing with a coherent manipulation \cite{Hirayama01,qd4}.

\begin{figure}[t]
\begin{center}    
\centering
\includegraphics[width=0.5\textwidth]{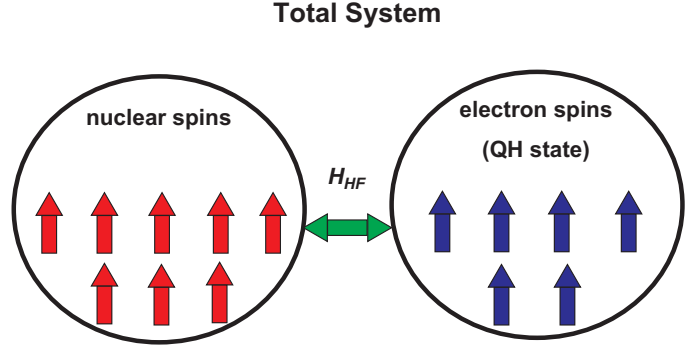}
\end{center}
\caption{A schematic illustration of ``hybrid'' quantum Hall (QH) systems. They consist of electron spins in the QH state and a large ensemble of nuclear spins in host crystals.
They are coupled through the hyperfine interaction $H_{\text{HF}}$, and therefore, the total system can be regarded as a hybrid QH systems. }
\label{qhsystemreservoir} 
\end{figure}

In this paper, for the first step of studying the 
electron-nuclear spin dynamics in the QH systems as  hybrid quantum systems (the schematic illustration is presented in FIG. \ref{qhsystemreservoir}), 
 we theoretically study the interaction between the nuclear spins and the linear dispersing NG mode associated with the U(1) spin rotational symmetry breaking.
 To make a detailed analysis and  clearly understand its physics behind, we focus on the CAF phase in the $\nu=2$ bilayer QH system.
We show that the NG mode couples with collective nuclear spins through the hyperfine interaction in the long-wavelength limit.  
As a result, such an interaction is effectively described by a Dicke model \cite{superradiance1,superradiance2,carmichaeltxb,agarwaltxb} with a
 continuous-mode,   which has been extensively studied in the quantum optics.
It is interesting that the interaction between nuclear spins and the linear dispersing NG mode 
mediated by the hyperfine interaction
in the QH systems can be described by the same model as the two-level atomic system coupled with photonic modes. 

Our analysis is not only valid for the nuclear spin-NG mode interaction in the CAF phase, but also for other QH systems where 
a linear dispersing NG mode is present due to the U(1) spin rotational symmetry breaking. 
For example, the skyrmion crystal formation in the monolayer QH system in the vicinity of the total filling factor $\nu=1$  \cite{cote1} may belong to this category,  
where a linear dispersing spin wave emerges and is expected to enhance the nuclear spin relaxation rate.

This paper is organized as follows. In Sec. \ref{sec2:hyperfine}, we analyze the hyperfine interaction in the QH systems. 
Based on this analysis, in Sec. \ref{sec3:Dickemodel}
we derive the Dicke model as the effective Hamiltonian describing the interaction between the nuclear spins and 
 the linear dispersing NG mode through the hyperfine interaction. It is the main result of this paper.

\section{Hyperfine Interaction in the QH systems}\label{sec2:hyperfine}
We first analyze the hyperfine interaction in QH systems  to derive the interaction Hamiltonian 
between the nuclear spins and the NG mode. As shown in FIG. \ref{qhsystemreservoir},
the QH systems consist of many-body nuclei and many-body electrons in the two-dimensional $xy$ plane. 
We may assume that all electrons are in  the $s$-type conduction band.
A high magnetic field is applied perpendicular to the plane, $\boldsymbol{B}=(0,0,-B_\perp)$ with $B_\perp>0$.

The interaction between nuclear and electron spins in the QH state is described by the contact hyperfine interaction \cite{hyperfine1,hyperfine2,hyperfine3},
\begin{align}
H_{\text{HF}}&=\frac{2\mu_0 \gamma_e \gamma_n \hbar^2}{3}
\sum_{i=1}^N {\boldsymbol{I}}_i
\cdot\sum_{j=1}^{N_e}|u(\boldsymbol{x}_j,z_j)|^2 {\boldsymbol{S}}_j 
\delta (\boldsymbol{X}_i-\boldsymbol{x}_j,Z_i-z_j)\notag\\
&=\frac{2\mu_0 \gamma_e \gamma_n \hbar^2}{3}
\sum_{i=1}^N  |u(\boldsymbol{X}_i,Z_i)|^2{\boldsymbol{I}}_i \cdot{\boldsymbol{S}}(\boldsymbol{X}_i,Z_i), \label{hyperfine1}
\end{align}
where ${\boldsymbol{S}}_j$ is the  electron spin at $(\boldsymbol{x}_j,z_j)$, 
${\boldsymbol{I}}_i$  the  nuclear spin at $(\boldsymbol{X}_i,Z_i)$ assuming spin 1/2,
and ${\boldsymbol{S}}(\boldsymbol{X}_i,Z_i)$ the three dimensional electron spin density.
At sufficiently low temperature electrons are confined within the lowest energy level,
and hence the motion of electrons along the $z$ direction is frozen.
We may approximate the quantum well by the system where 
electrons are located at the center of the well, that is, $z_j=z_0$ for all $j$,
and interact with nuclear spins at $Z_i=z_0$ for all $i$. 
Then we may set  ${\boldsymbol{S}}(\boldsymbol{X}_i,Z_i)\simeq{\boldsymbol{S}}(\boldsymbol{X}_i)L_z^{-1}$
for $-L_z/2<Z_i<L_z/2$  with  $L_z$ the width of quantum well
and ${\boldsymbol{S}}(\boldsymbol{X}_i)$ the two dimensional electron spin density.
The quantities  $\gamma_e$ $(\gamma_n)$, $\mu_0$,   $u(\boldsymbol{X}_i,z_0)$, $N$, and $N_e$
are the gyromagnetic ratio for  electron (nucleon), the permeability of vacuum magnetic constant,  the Bloch amplitude at $(\boldsymbol{X}_i,z_0)$,
 the total number of  polarized nuclear spins,  and  the total electron number, respectively. 
Since the Bloch amplitude is 
a periodic function with respect to the nuclear spin separation, we can set $|u(\boldsymbol{X}_i,z_0)|^2=\eta=$const. 
Then the hyperfine interaction \eqref{hyperfine1} is rewritten as
\begin{align}
&H_{\text{HF}}
=A\sum_{i=1}^N {\boldsymbol{I}}_i \cdot {\boldsymbol{S}}(\boldsymbol{X}_i),\qquad\text{with}\quad
A=\frac{2\mu_0 \gamma_e \gamma_n \hbar^2\eta}{3L_z}. \label{hyperfine2}
\end{align}
The values of $\eta$ for Ga and As are given by $\eta_{\text{Ga}}=2.7\times10^3$ and  $\eta_{\text{As}}=4.5\times10^3$, respectively \cite{hyperfine0}. 
Here we take  $\eta=10^3.$

The hyperfine coupling is weak compared with  the Landau-level energy as well as the thermal energy.
It is reasonable to assume that the electronic 
ground state is unchanged by the  hyperfine interaction, and to  replace the electron spin density in Eq. \eqref{hyperfine2} by the classical spin density 
${\boldsymbol{S}}^{\text{cl}}(\boldsymbol{X_i})=\langle \phi_{\text{QH}}|
{\boldsymbol{S}}(\boldsymbol{X_i})
|\phi_{\text{QH}}\rangle$, with $|\phi_{\text{QH}}\rangle$ denoting the QH state. 
We express the hyperfine interaction in terms
of normalized spin density defined by ${\boldsymbol{\mathcal{S}}}^{\text{cl}}(\boldsymbol{X_i})=\rho_\Phi^{-1}{\boldsymbol{S}}^{\text{cl}}(\boldsymbol{X_i})$, 
where $\rho_\Phi=\rho_0/\nu=1/2\pi l_B^2$ is the density of states of the Landau sites with $\rho_0$ the total electron density, $\nu$ the total Landau filling factor, and $l_B$ the magnetic length.
Eq. \eqref{hyperfine2} becomes
\begin{align}
H_{\text{HF}}=\tilde{g}\sum_{i=1}^N {\boldsymbol{I}}_i \cdot
{\boldsymbol{\mathcal{S}}}^{\text{cl}}(\boldsymbol{X}_i), \label{hyperfine3}
\end{align}
where $\tilde{g}=A\rho_\Phi.$
The Hamiltonian \eqref{hyperfine3} describes the  hyperfine interaction between the nuclear spins 
and the electron spins in the QH state. 
The order of the coupling $\tilde{g}$ in  \eqref{hyperfine3} is estimated by setting $\mu_0=4\pi\times10^{-7}$ N$\cdot$A$^{-2}$,  $\gamma_e=1.761\times10^{11}$  rad/s$\cdot$T,  
$\gamma_n=10\times10^{7}$  rad/s$\cdot$T,  $\rho_0=1\times10^{15}$ m$^{-2}$, $L_z=10^{-8}$ m,
$\eta=10^3$, and $\hbar=1.0546\times10^{-34}$ J$\cdot$s/rad, and
we have $\tilde{g}/\hbar\sim100$ rad$\cdot$kHz/T. It is much smaller compared with the Larmor frequency $\omega_{\text{s}}\sim10$ rad$\cdot$MHz/T.

\section{Dicke Model in the QH systems}\label{sec3:Dickemodel}
We next derive the interaction Hamiltonian between the nuclear spins and the NG mode.
We show that collective nuclear spins couple with the NG mode in the long-wavelength limit, 
and furthermore that  it is effectively described by the Dicke model.

\subsection{Electron Spin Configuration and Effective Hamiltonian for the NG mode }\label{subsec31:nginformation}
When the U(1) spin rotational symmetry is spontaneously broken around the $z$-axis, 
 the in-plane component of the classical electron spin density is expressed as 
\begin{align}
{\mathcal{S}}^{\text{cl},x}(\boldsymbol{x})=S\cos(\vartheta_0+\delta\vartheta(\boldsymbol{x})), \qquad
{\mathcal{S}}^{\text{cl},y}(\boldsymbol{x})=S\sin(\vartheta_0+\delta\vartheta(\boldsymbol{x})), \label{inplanespinconfiguration}
\end{align}
where $S$  is a constant in the range $0< |S|<1$.
In the ground state the electron spins are in a spatially homogeneous configuration with a fixed orientation angle $\vartheta_0$ of the in-plane spin component.
The fluctuation field $\delta\vartheta(\boldsymbol{x})$ is the associated NG mode. 
For the $z$-component, it is assumed that the fluctuation of the electron spin density is negligible,
because it is gapped.

We expand the above spin densities in terms of  $\delta\vartheta(\boldsymbol{x})$ up to the first order and 
 substitute it to the hyperfine interaction \eqref{hyperfine3}. 
The zeroth order terms with respect to $\delta\vartheta(\boldsymbol{x})$
are 
\begin{align}
H_{\text{HF}}^{(0)}=
\left(\tilde{g}S\sum_i ( {I}^x_i\cos\vartheta_0+ {I}^y_i\sin\vartheta_0)\right)+\tilde{g}\sum_i{I}_i^z\mathcal{S}^g_z, \label{hyperfine4}
\end{align}
where $\mathcal{S}^g_z$ is the ground-state expectation value of the $z$ component of the normalized electron spin density satisfying $0<|\mathcal{S}^g_z|<1$.
The first term   describes the in-plane magnetic field induced by the hyperfine interaction, which nuclear spins experience, while 
 the second term generates the Knight shift $K_{\text{s}}\omega_{\text{s}}$.
These two terms have the same order of magnitude as the coupling $\tilde{g}/\hbar\sim100$ rad$\cdot$kHz/T, 
which was mentioned in the previous section, i.e., $K_{\text{s}}\omega_{\text{s}}\sim100$ rad$\cdot$kHz/T. 
Here we note that this value is comparable with the experimental result  reported in \cite{Kumadaetal}, i.e., $(K_{\text{s}}\omega_{\text{s}})^{\text{Exp}}\sim10$ rad$\cdot$kHz.
We can drop these terms because they are negligible compared with the nuclear-spin Larmor frequency,
which is $\omega_{\text{s}}\sim 10$ rad$\cdot$MHz/T.
 
\begin{figure}[t]
\begin{center}    
\centering
\includegraphics[width=0.7\textwidth]{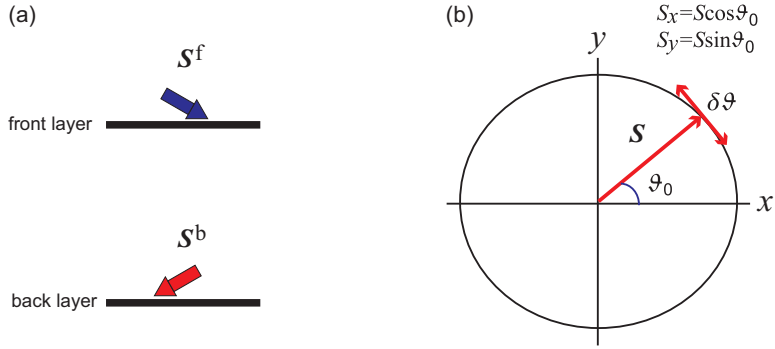}
\end{center}
\caption{Electron-spin configuration in the CAF phase. (a) Electron spins have ferromagnetic correlation in each layer, while antiferromagnetic correlation between the two layers. Consequently, electron spins are aligned (canted) coherently. 
(b) The electron spin for in-plane component in the front layer. We denote this plane as $xy$ plane whereas the $z$ axis indicate the direction perpendicular to the $xy$ plane. 
The electron spins are in a spatially  homogeneous configuration, characterized by the orientation angle $\vartheta_0.$ The fluctuation mode $\delta\vartheta$ is the NG mode.   }
\label{spinconfiguration} 
\end{figure}

 The first-order terms in  $\delta\vartheta(\boldsymbol{x})$
 describe the interaction between the nuclear spins and the NG mode,
\begin{align}
H_{\text{HF}}^{(1)}=
g\sum_i (- {I}^x_i\sin\vartheta_0+ {I}^y_i\cos\vartheta_0)\delta\vartheta(\boldsymbol{x}), 
\qquad \text{with} \qquad g=\tilde{g}S.
\label{hyperfine4-1}
\end{align}
To understand the situation more clearly,
we present an example, the CAF phase in the $\nu=2$ bilayer QH state. 
As presented in FIG. \ref{spinconfiguration}, 
electron spins have  ferromagnetic correlations in each layer, 
whereas they have antiferromagnetic correlations between the two layers. 
The electron spin configuration in the front (back) layer 
${\mathcal{S}}^{\text{f(b)}}_a$ $(a=x,y,z)$
is described as
\begin{align}
{\mathcal{S}}^{\text{f}}_x=-{\mathcal{S}}^{\text{b}}_{x}=S\cos\vartheta_0, \qquad
{\mathcal{S}}^{\text{f}}_y=-{\mathcal{S}}^{\text{b}}_{y}=S\sin\vartheta_0, \qquad
{\mathcal{S}}^{\text{f}}_z={\mathcal{S}}^{\text{b}}_{z}=\mathcal{S}_z^{\text{caf}},
\label{cafspinconfiguration}
\end{align}
where ${\mathcal{S}}^{\text{f(b)}}_{a}$ $(a=x,y)$ and ${\mathcal{S}}^{\text{f(b)}}_z$ are the in-plane and
 $z$ components of the electron spin in the front (back) layer, respectively. Here $S$ and $\mathcal{S}_z^{\text{caf}}$
satisfy the conditions $0<|S|<1$ and $0<|\mathcal{S}_z^{\text{caf}}|<1$.
As a result, spins are canted coherently.
By focusing  on the in-plane component, electron spins orient homogeneously characterized by an 
angle $\vartheta_0$. Although the ground state energy does not depend on   $\vartheta_0$, 
the ground state itself does, reflecting that the CAF state is the U(1) spin rotational symmetry broken state around $z$ axis.
The small fluctuation mode $\delta\vartheta$ is the corresponding NG mode. 
We present the explicit formula of \eqref{cafspinconfiguration} for the case of the CAF phase in Appendix \ref{appendixA}.

The intriguing feature of the NG mode is that it has a linear dispersion in the CAF phase.
We denote the Fourier transform of $\delta\vartheta(\boldsymbol{x})$ as $\delta\vartheta_{\boldsymbol{k}}$, 
\begin{align}
\delta\vartheta_{\boldsymbol{k}}=\int \frac{d^2x}{2\pi} e^{-i\boldsymbol{k}\boldsymbol{x}}\delta\vartheta(\boldsymbol{x}),\label{thetaft}
\end{align}
and introduce the canonical conjugate variable $\delta\sigma_{\boldsymbol{k}}$ satisfying 
$[\delta\sigma_{\boldsymbol{k}},\delta\vartheta^\dagger_{\boldsymbol{k}^\prime}]=i\delta(\boldsymbol{k}-\boldsymbol{k}^\prime)$.
By making the momentum expansion, the effective Hamiltonian for the NG mode has the following form,
\begin{align}
H_{\text{R}}=\int d^2k \left[
(a+b\boldsymbol{k}^2)\delta\sigma^\dagger_{\boldsymbol{k}}\delta\sigma_{\boldsymbol{k}}
+(c\boldsymbol{k}^2)\delta\vartheta^\dagger_{\boldsymbol{k}}\delta\vartheta_{\boldsymbol{k}}
\right], \label{abcsigmathetanghamiltonian}
\end{align}
where $a,$ $b$ and $c$ are positive constants. 
This Hamiltonian is diagonalized by introducing another set of canonical variables,
\begin{align}
r_{\boldsymbol{k}}=\frac{1}{\sqrt{2}}\left(
\sqrt{G_{\boldsymbol{k}}}\delta{\sigma_{\boldsymbol{k}}}+i\frac{1}{\sqrt{G_{\boldsymbol{k}}}}
{\vartheta_{\boldsymbol{k}}}
\right), \qquad
r^\dagger_{\boldsymbol{k}}=\frac{1}{\sqrt{2}}\left(
\sqrt{G_{\boldsymbol{k}}}\delta{\sigma}_{\boldsymbol{k}}^\dagger-i\frac{1}{\sqrt{G_{\boldsymbol{k}}}}
{\vartheta^\dagger_{\boldsymbol{k}}}
\right),\label{ngcranoperator}
\end{align}
satisfying
$\left[ r_{\boldsymbol{k}}, r^\dagger_{\boldsymbol{k}^\prime}\right]=\delta(\boldsymbol{k}-\boldsymbol{k}^\prime),$
where 
\begin{align}
G_{\boldsymbol{k}}=\left(\frac{a+b\boldsymbol{k}^2}{c\boldsymbol{k}^2} \right)^{\frac{1}{2}}.
\label{relationabcG}
\end{align}
Now
the effective Hamiltonian \eqref{abcsigmathetanghamiltonian} is diagonalized as
\begin{align}
{H}_{\text{R}}&=\int d^2 k E_{\boldsymbol{k}}r^\dagger_{\boldsymbol{k}}r_{\boldsymbol{k}}, \qquad
\text{with} \qquad
E_{\boldsymbol{k}}=2\sqrt{c\boldsymbol{k}^2(a+b\boldsymbol{k}^2)}\approx
2\sqrt{ac}|\boldsymbol{k}|,
\label{rEffecHamiltonian}
\end{align}
with $E_{\boldsymbol{k}}$ a linear dispersion relation for the NG mode.
We present the explicit formulas of 
Eqs. \eqref{abcsigmathetanghamiltonian}, \eqref{relationabcG}, and \eqref{rEffecHamiltonian} 
for the case of the CAF phase in Appendix \ref{appendixB}: See Eqs. 
\eqref{nghamiltonian1}, \eqref{gdefinition}, and \eqref{nglineardispersion1}, respectively.

\subsection{Dicke Model }\label{subsec32:Dickemodel}
We now describe the hyperfine interaction Hamiltonian \eqref{hyperfine4-1} in terms of 
the NG mode $r_{\boldsymbol{k}}$, $r^\dagger_{\boldsymbol{k}}$ and the nuclear spins 
${I}^{x,y}_i$. Using Eqs. \eqref{thetaft} and   \eqref{ngcranoperator}, we obtain
\begin{align}
H_{\text{HF}}&=\frac{g}{2}\sum_{i=1}^N (\tilde{I}_i^++\tilde{I}_i^-)
\int \frac{d^2k}{2\pi}\left(\kappa_{\boldsymbol{X}_i,\boldsymbol{k}}r_{\boldsymbol{k}}
+\kappa^\ast_{\boldsymbol{X}_i,\boldsymbol{k}}r^\dagger_{\boldsymbol{k}})
\right), \notag\\
\kappa_{\boldsymbol{X}_i,\boldsymbol{k}}&=\sqrt{\frac{G_{\boldsymbol{k}}}{2}}e^{i(\boldsymbol{k}\boldsymbol{X}_i-\frac{\pi}{2})}, \quad 
\kappa^\ast_{\boldsymbol{X}_i,\boldsymbol{k}}=\sqrt{\frac{G_{\boldsymbol{k}}}{2}}e^{-i(\boldsymbol{k}\boldsymbol{X}_i-\frac{\pi}{2})}.
\label{hyperfine5}
\end{align}
where  $\tilde{I}^{\pm}=e^{\mp i(\vartheta_0+\pi/2)}{I}^{\pm}$ $(I^{\pm}=I^x\pm iI^y)$ is the rotated in-plane nuclear spin. In the rest of this paper  we just write  $\tilde{I}^{\pm}$ as  ${I}^{\pm}$. 
To derive \eqref{hyperfine5}, we used the relations  $G_{\boldsymbol{k}}=G_{-\boldsymbol{k}}$, $\delta{\sigma}^\dagger_{\boldsymbol{k}}=\delta{\sigma}_{-\boldsymbol{k}}$ and
$\vartheta^\dagger_{\boldsymbol{k}}=\vartheta_{-\boldsymbol{k}}.$ They follow from the fact that
$G_{\boldsymbol{k}}$ are even function of $\boldsymbol{k},$ and
 $\delta{\sigma}(\boldsymbol{x})=\int(d^2k/2\pi)e^{-i\boldsymbol{k}\boldsymbol{x}}\delta{\sigma}^\dagger_{\boldsymbol{k}}$ and $\delta\vartheta(\boldsymbol{x})$ are real fields.
Using the rotating-wave  approximation for \eqref{hyperfine5},
we obtain
\begin{align}
H_{\text{SR}}&=\frac{g}{2}\sum_{i=1}^N (I^+_i R_i+ I^-_i R^\dagger_i), \notag\\
R_i&=\int \frac{d^2k}{2\pi}\kappa_{\boldsymbol{X}_i,\boldsymbol{k}}r_{\boldsymbol{k}}, \quad
R_i^\dagger=\int \frac{d^2k}{2\pi}\kappa^\ast_{\boldsymbol{X}_i,\boldsymbol{k}}r^\dagger_{\boldsymbol{k}}.
\label{hyperfine6}
\end{align}
Furthermore, the $i$ dependence disappears from $R_i$ and $R_i^\dagger$
in the long wave-length limit $e^{i\boldsymbol{k}\boldsymbol{X}_i}\approx1$.
We may rewrite (\ref{hyperfine6}) as
\begin{align}
H_{\text{SR}}&=\frac{g}{2}\sum_{i=1}^N (I^+_i R+ I^-_i R^\dagger)
=\frac{g}{2} (J^+ R+ J^- R^\dagger), \notag\\ 
R&=\int \frac{d^2k}{2\pi}\kappa_{\boldsymbol{k}}r_{\boldsymbol{k}}, \quad
R^\dagger=\int \frac{d^2k}{2\pi}\kappa^\ast_{\boldsymbol{k}}r^\dagger_{\boldsymbol{k}}, \notag\\
\kappa_{\boldsymbol{k}}&=\sqrt{\frac{G_{\boldsymbol{k}}}{2}}e^{-i\frac{\pi}{2}}, \quad 
\kappa^\ast_{\boldsymbol{X}_i,\boldsymbol{k}}=\sqrt{\frac{G_{\boldsymbol{k}}}{2}}e^{i\frac{\pi}{2}}.
\label{hyperfine7}
\end{align}
Indeed, as the NG mode has a long wavelength,
the approximation $e^{i\boldsymbol{k}\boldsymbol{X}_i}\approx1$ is valid in this system.
For instance, in the case of the CAF phase the wavelength of the NG mode becomes $\lambda_{\text{s}}\sim10^7$\AA\ 
for $E_{\boldsymbol{k}}=\hbar\omega_{\text{s}}$ with $\omega_{\text{s}}\sim10$ rad$\cdot$MHz/T,
where $E_{\boldsymbol{k}}=\gamma|\boldsymbol{k}|$ $(\gamma=2\sqrt{ac}>0)$ is the linear dispersion relation for the NG mode. 
The value of $\lambda_{\text{s}}$ is about the same as the sample size $L\sim100\mu$m (see Eq. \eqref{ngmodewavelength}). 
Thus it is a good approximation at this energy scale. 
If the dispersion for the NG mode were quadratic with the same coefficient as the linear one, 
the wavelength at $E_{\boldsymbol{k}}=\hbar\omega_{\text{s}}$ would be around $\lambda_{\text{s}}\sim10^3$ \AA,
which is much smaller than the sample size. 
Such a case, the long-wavelength limit is not a good approximation 
so that the interaction Hamiltonian $H_{\text{SR}}$ cannot be expressed 
in terms of the NG mode and the collective spin.

As a result, the interaction between the nuclear spins and the linear dispersing NG mode 
is described as 
an interaction between the  collective nuclear spin $J=\sum_i I_i$ and the NG mode $R$ with the coupling constant $g$ in the long-wavelength limit.    

The other relevant terms are 
the nuclear-spin Hamiltonian $H_{\text{S}}$ describing the Larmor precision, 
\begin{align}
H_{\text{S}}&=-\hbar\gamma_n B_z\sum_{i=1}^N {I}_{i}^z=\hbar\omega_{\text{s}}\sum_{i=1}^N 
{I}_{i}^z=\hbar \omega_{\text{s}}{J}^{z}. \label{Larmorprecision}
\end{align}
The total effective Hamiltonian  $H=H_{\text{S}}+H_{\text{R}}+H_{\text{SR}}$ is given by
\begin{align}
H=\hbar \omega_{\text{s}}{J}^{z}+\int d^2 k E_{\boldsymbol{k}}r^\dagger_{\boldsymbol{k}}r_{\boldsymbol{k}}
+\frac{g}{2} (J^+ R+ J^- R^\dagger).
\label{QHdicke}
\end{align}
Consequently, the interaction between the nuclear spins and the linear dispersing NG mode mediated by the hyperfine interaction is described effectively by 
 the Dicke model \cite{superradiance1,superradiance2,carmichaeltxb,agarwaltxb} with a continuous-mode,
where the collective spin operator $J^a$ $(a=x,y,z)$ with its magnitude $N/2$ interacts with the NG mode with the coupling ${g}$. 
It is interesting that the nuclear spin-NG mode interaction mediated by the hyperfine interaction in the QH systems can be described by the analogue model 
as the two-level atomic systems surrounded by the electromagnetic field in the vacuum:  Nuclear-spins 1/2 correspond to the 
two-level atoms, whereas the NG mode corresponds to the electromagnetic field in the vacuum \cite{agarwaltxb}.
  For an explicit derivation of this Dicke model in the case of the CAF phase, 
see Appendix \ref{appendixC}.

\section{Conclusion}\label{sec6:theoryexp}
In this paper, we have presented the theoretical studies on the interaction between the nuclear spins 
 and the linear dispersing NG
mode due to the spin U(1) rotational symmetry breaking in the QH systems mediated by the hyperfine interaction. 
Since  the NG mode has a long wavelength, the nuclear spins  couple collectively with
 the NG mode. Consequently, the nuclear spin-NG mode interaction can be described effectively by the 
Dicke model with a continuous-mode.
 This physics could be captured by regarding the QH systems as  hybrid quantum systems comprise of electron and nuclear spins.  
 In this paper, we  focused on  the CAF phase in $\nu=2$ bilayer QH systems and demonstrated that the interaction between the nuclear spins and the NG mode is described by the above Dicke model.
This Dicke model must be also applied to QH systems with the linear dispersing NG mode due to the spin U(1) rotational symmetry breaking in general, for instance, 
 the nuclear spin-NG mode interaction in the vicinity of $\nu=1$ monolayer QH systems in the presence of the skyrmion crystal formation.
 
 We would like to emphasize that the Dicke model derived in this paper can lead to  new directions
 for the study of QH physics, which cannot be obtained solely from the perspectives of the solid state physics. 
Combining the perspectives of  solid state physics with these from quantum optics,  we may reveal rich and new phenomena 
in the QH systems.
For instance,  the realization of phenomena similar to those in the quantum optical systems, for example, 
superradiance in terms of nuclear spins in the QH systems might be one possibility. The study of the spin-boson dynamics inherent to the QH system might be another interesting direction.

\acknowledgements
Y.~Hama thanks Naoto Nagosa, Makoto Yamaguchi,  and Franco Nori for fruitful discussion and comments. 
We thank Adam Miranowicz  for carefully reading the manuscript. 
This work was supported in part by RIKEN Special Postdoctoral 
Researcher Program (Y.~Hama), Interdepartmental Doctoral Degree Program for
Multi-dimensional Materials Science Leaders (Tohoku University) (M.~H.~Fauzi and Y.~Hirayama), 
JSPS KAKENHI Grant Number 25220601 and MEXT KAKENHI Grant number 15H05870 (K.~Nemoto),
ERATO Nuclear Spin Electronics Project (JST) and MEXT KAKENHI Grant Numbers 15H05867
and 26287059 (Y.~Hirayama),
and Grants-in-Aid for Scientific Research from the Ministry of Education, Science, Sports and Culture 
(Z.~F.~Ezawa).

\appendix

\section{Order Parameters in the CAF phase}\label{appendixA}

We present the derivation of the Dicke model by constructing {the Hamiltonian $H_{\text{R}}$ for the NG mode and the interaction Hamiltonian $H_{\text{SR}}$ between the nuclear spin and the NG mode in the case of the CAF phase.
To derive the Dicke model, we make a concise review of the  effective Hamiltonian density for the ground state and the NG modes 
in bilayer QH systems \cite{Ezawa:2005xi,yhamaetal,Ezawa:2013ae}.  
We start with the  discussion on the phase structure as well as the spin density configuration, 
and the associated NG modes at $\nu=2.$ 
Then we present  the effective Hamiltonian density for the linear dispersing NG mode in the CAF phase.

In the bilayer QH systems electrons possess the four internal degrees of freedom, the spin and the layer (pseudospin). 
We denote the two layers as the ``front" and ``back" layers. The electron field operator in the bilayer QH systems 
is represented as $\Psi(\boldsymbol{x})=(\psi_{\uparrow\text{f}}(\boldsymbol{x}),\psi_{\downarrow\text{f}}(\boldsymbol{x}),\psi_{\uparrow\text{b}}(\boldsymbol{x}),\psi_{\downarrow\text{b}}(\boldsymbol{x}))$. 
The physical operators  in this system are expressed in terms of the following sixteen operators, 
the density operator and SU(4) isospin operators. In terms of the electron field operator $\Psi(\boldsymbol{x})$,
they are given by
\begin{align}
\rho (\boldsymbol{x})& =\Psi ^{\dagger }(\boldsymbol{x})\Psi (\boldsymbol{x}),  \quad
S_{a}(\boldsymbol{x}) =\frac{1}{2}\Psi ^{\dagger }(\boldsymbol{x})\tau _{a}^{\text{spin}}
\Psi (\boldsymbol{x}),  \notag \\
P_{a}(\boldsymbol{x})& =\frac{1}{2}\Psi ^{\dagger }(\boldsymbol{x})\tau _{a}^{\text{ppin}}
\Psi (\boldsymbol{x}),  \quad 
R_{ab}(\boldsymbol{x}) =\frac{1}{2}\Psi ^{\dagger }(\boldsymbol{x})\tau _{a}^{\text{spin}}
\tau _{b}^{\text{ppin}}\Psi (\boldsymbol{x}),  \label{physicaloperator1}
\end{align}
where $a,b=x,y,z$ and
\begin{align}
\tau _{a}^{\text{spin}}=\left( 
\begin{array}{cc}
\tau _{a} & 0 \\ 
0 & \tau _{a}
\end{array}
\right) ,  \quad
\tau _{x}^{\text{ppin}} =  \left( 
\begin{array}{cc}
0 & \boldsymbol{1}_{2} \\ 
\boldsymbol{1}_{2} & 0
\end{array}
\right) ,\quad 
\tau _{y}^{\text{ppin}}=\left( 
\begin{array}{cc}
0 & -i\boldsymbol{1}_{2} \\ 
i\boldsymbol{1}_{2} & 0
\end{array}
\right) , \quad
\tau _{z}^{\text{ppin}} =   \left( 
\begin{array}{cc}
\boldsymbol{1}_{2} & 0 \\ 
0 & -\boldsymbol{1}_{2}
\end{array}
\right) ,  \label{su4base}
\end{align}
with $\tau_a$ denoting the Pauli matrices. The operator $\rho (\boldsymbol{x})$ is the density operator,
while the SU(4) isospin density operator $S_{a}(\boldsymbol{x})$, $P_{a}(\boldsymbol{x})$ and $R_{ab}(\boldsymbol{x})$ 
represent the spin density operator, the pseudospin density operator, and the $R$-spin density operator, respectively.
On the other hand, the total Hamiltonian in this system is given by \cite{Ezawa:2013ae},  $H=H_{\text{K}}+H_{\text{C}}+H_{\text{Z}}+H_{\text{PZ}}$, 
where $H_{\text{K}}$ is the kinetic term which generates the Landau level, 
$H_{\text{C}}$ the Coulomb interaction term, $H_{\text{Z}}$ the Zeeman interaction term, 
and $H_{\text{PZ}}$ the pseudo-Zeeman term composed of the tunnelling interaction term and the bias term,
which represents the creation of the density-imbalanced configuration between the two layers.
The Coulomb interaction is decomposed into the form $H_{\text{C}}=H_{\text{C}}^++H_{\text{C}}^-,$
where $H_{\text{C}}^{+(-)}$ represents the SU(4) invariant (non-invariant) term.

We analyze the classical Hamiltonian density of  $H=\int d^2x \mathcal{H} (\boldsymbol{x})$
by performing the lowest Landau level projection. We denote the generic lowest Landau level state as $| \mathfrak{S}\rangle$. We derive the form of $\mathcal{H}^{\text{cl}} (\boldsymbol{x})=\langle\mathfrak{S} | \mathcal{H} (\boldsymbol{x})
|\mathfrak{S} \rangle$, which are represented by the classical density operators
$\rho^{\text{cl}} (\boldsymbol{x})=\langle\mathfrak{S} |\rho (\boldsymbol{x})
|\mathfrak{S} \rangle$, $S^{\text{cl}}_a (\boldsymbol{x})=\langle\mathfrak{S} |S_a (\boldsymbol{x})
|\mathfrak{S} \rangle$, $P^{\text{cl}}_a (\boldsymbol{x})=\langle\mathfrak{S} |P_a (\boldsymbol{x})
|\mathfrak{S} \rangle$,  and $R^{\text{cl}}_{ab} (\boldsymbol{x})=\langle\mathfrak{S} |R_{ab} (\boldsymbol{x})
|\mathfrak{S} \rangle$. Actually we express $\mathcal{H}^{\text{cl}}$ in terms of the normalized SU(4) 
operators defined by  $S^{\text{cl}}_a (\boldsymbol{x})=\rho_\Phi \mathcal{S}^{\text{cl}}_a (\boldsymbol{x})$,
 $P^{\text{cl}}_a (\boldsymbol{x})=\rho_\Phi \mathcal{P}^{\text{cl}}_a (\boldsymbol{x})$, and  
$R^{\text{cl}}_{ab} (\boldsymbol{x})=\rho_\Phi \mathcal{R}^{\text{cl}}_{ab} (\boldsymbol{x})$,
by setting $\rho^{\text{cl}}  (\boldsymbol{x})=\rho_0$  due to the incompressibility of the QH state.

In the $\nu=2$ bilayer QH system the SU(4) order parameters, which are the expectation values of the normalized SU(4) operators in the ground state, are given by \cite{Ezawa:2005xi}  
\begin{align} 
&\mathcal{S}_{z}^{0} =-\frac{\Delta _{\text{Z}}}{\Delta _{0}}(1-\alpha ^{2})
\sqrt{1-\beta ^{2}},  \notag\\
&\mathcal{P}_{x}^{0} =\frac{\Delta _{\text{SAS}}}{\Delta _{0}}\alpha ^{2}
\sqrt{1-\beta ^{2}},\ \ \mathcal{P}_{z}^{0}=\frac{\Delta _{\text{SAS}}}
{\Delta _{0}}\alpha ^{2}\beta=\sigma_0 ,  \notag \\
&\mathcal{R}_{xx}^{0}+i\mathcal{R}_{yx}^{0} =-\frac{\Delta _{\text{SAS}}}
{\Delta _{0}}\alpha \sqrt{1-\alpha ^{2}}\beta e^{-i\omega},  \notag\\
&\mathcal{R}_{yy}^{0}+i\mathcal{R}_{xy}^{0}=\frac{\Delta _{\text{Z}}}{\Delta _{0}}
\alpha \sqrt{1-\alpha ^{2}}\sqrt{1-\beta ^{2}}e^{i\omega},  \notag\\
&\mathcal{R}_{xz}^{0}+i\mathcal{R}_{yz}^{0}=\frac{\Delta _{\text{SAS}}}
{\Delta _{0}}\alpha \sqrt{1-\alpha ^{2}}\sqrt{1-\beta ^{2}}e^{-i\omega},
\label{orderparameter1}
\end{align}
with
\begin{equation}
\Delta _{0}\equiv \sqrt{\Delta _{\text{SAS}}^{2}\alpha ^{2}+\Delta _{\text{Z}%
}^{2}\left( 1-\alpha ^{2}\right) \left( 1-\beta ^{2}\right) },
\end{equation}%
where  $\Delta _{\text{Z}}$ is the Zeeman gap and $\Delta _{\text{SAS}}$ the tunneling gap; 
$\alpha$, $\beta$ ($|\alpha|,|\beta|\leq1$) and $\omega$ are real parameters. 
The quantity $\sigma_0$ is the imbalanced parameter, representing the density difference between 
the front and back layers, and  defined by $\sigma_0=(\rho_0^{\text{f}}-\rho_0^{\text{b}})/(\rho_0^{\text{f}}+\rho_0^{\text{b}})$,
with $\rho_0^{\text{f(b)}}$ the electron density in the front (back) layer.
Here the parameters
$\alpha$ and $\beta$ are determined by minimizing the classical Hamiltonian given by \eqref{su4effectivehamiltonian1},
where the normalized isospin densities are in spatially homogeneous configurations.
They satisfy the condition $(\mathcal{S}_{a}^{0})^2+ (\mathcal{P}_{a}^{0})^2+ (\mathcal{R}_{ba}^{0})^2=1. $
We demonstrate later that the parameter $\omega$ is associated with the NG mode in the CAF phase: 
See (\ref{NGmode4}).

It was shown \cite{Ezawa:2005xi} that  for $\alpha=0,$ the ground state is the ferromagnetic phase, 
where only the spin is polarized as $\mathcal{S}_{z}^{0}=1$. 
On the other hand, for $\alpha=1$ the ground state is the spin-singlet phase, 
where only the pseudospin is polarized as $\mathcal{P}_{z}^{0} =\sigma_0$ and
$\mathcal{P}_{x}^{0} =\sqrt{1-\sigma^2_0}$.
The CAF phase is realized for $0<\alpha<1$.

From  $2\mathcal{S}^{\text{f}}_{a}=\mathcal{S}^0_{a}+\mathcal{R}^0_{az}$ and 
$2\mathcal{S}^{\text{b}}_{a}=\mathcal{S}^0_{a}-\mathcal{R}^0_{az}$, 
the relations between the spin densities in the front and back layers are
\begin{align}
\mathcal{S}^{\text{f}}_{x}&=-\mathcal{S}^{\text{b}}_{x}=\frac{1}{2}\frac{\Delta _{\text{SAS}}}
{\Delta _{0}}\alpha \sqrt{1-\alpha ^{2}}\sqrt{1-\beta ^{2}}\cos\omega,   \notag\\
\mathcal{S}^{\text{f}}_{y}&=-\mathcal{S}^{\text{b}}_{y}=-\frac{1}{2}\frac{\Delta _{\text{SAS}}}
{\Delta _{0}}\alpha \sqrt{1-\alpha ^{2}}\sqrt{1-\beta ^{2}}\sin\omega,   \notag\\
\mathcal{S}^{\text{f}}_{z}&=\mathcal{S}^{\text{b}}_{z}=\frac{1}{2}\mathcal{S}^0_{z}.   \label{cafspincomponent1}
\end{align}
From  the above equation, we see that in the CAF phase the 
antiferromagnetic correlation is built up  between the two layers. Here the angle  $\omega$ describes the orientation angle of the in-plane spin component. The order parameters \eqref{orderparameter1} 
  are obtained from the ones with $\omega=0$ by the spin rotation 
exp$[iT_{z0}\omega]$.
Furthermore, the  Hamiltonian density \eqref{su4effectivehamiltonian1} is invariant under this rotation.
Thus, the CAF phase is the $\text{U}_{T_{z0}}(1)$ spin rotational symmetry broken state.

At $\nu =2$, four complex NG modes emerge due to the symmetry breaking
pattern SU(4)$\rightarrow $U(1)$\times $SU(2)$\times $SU(2). All the NG
modes get gapped due to the presence of the Coulomb interactions, the Zeeman
and pseudo-Zeeman terms. We analyze the system in the limit $\Delta _{\text{SAS}}\rightarrow 0$, 
where only one NG mode responsible to the interlayer phase coherence becomes gapless. 
In this limit the
order parameters \eqref{orderparameter1} are reduced to 
\begin{align}
\mathcal{S}_{z}^{0}& =|\sigma _{0}|-1,\quad \mathcal{P}_{z}^{0}=\sigma _{0},
\notag \\
\mathcal{R}_{yy}^{0}& =-\rm{sgn}(\sigma_0)\mathcal{R}_{xx}^{0}=\sqrt{|\sigma _{0}|(1-|\sigma
_{0}|)}\cos \omega ,\quad \mathcal{R}_{xy}^{0}=\rm{sgn}(\sigma_0)\mathcal{R}_{yx}^{0}
=\sqrt{|\sigma _{0}|(1-|\sigma _{0}|)}\sin \omega ,  \label{orderparameter2}
\end{align}
with the relations
\begin{equation}
\frac{\Delta _{\text{Z}}}{\Delta _{0}}
\sqrt{1-\beta ^{2}}=1,\quad \frac{\Delta _{\text{SAS}}}
{\Delta _{0}}=1,\quad \alpha ^{2}=|\sigma _{0}|,\quad \sqrt{1-\beta ^{2}}=0,\quad 
\beta=\rm{sgn}\sigma _{0},
\end{equation}
which are obtained in the limit  $\Delta _{\text{SAS}}\rightarrow 0$.
We consider the case $\sigma _{0}>0$ explicitly, while the case $\sigma _{0}<0$ is
similarly discussed. The order parameters (\ref{orderparameter2}) imply that
the SU(4) isospins are given by 
\begin{align}
& \mathcal{S}_{z}(\boldsymbol{x})=\sigma (\boldsymbol{x})-1,\quad \mathcal{P}_{z}(\boldsymbol{x})=\sigma (\boldsymbol{x}),  \notag \\
& \mathcal{R}_{yy}(\boldsymbol{x})=-\mathcal{R}_{xx}(\boldsymbol{x})
=\sqrt{\sigma (\boldsymbol{x})(1-\sigma (\boldsymbol{x}))}\cos \vartheta (\boldsymbol{x}),\quad 
\mathcal{R}_{xy}(\boldsymbol{x})=\mathcal{R}_{yx}(\boldsymbol{x})=-\sqrt{\sigma (\boldsymbol{x})(1-\sigma (\boldsymbol{x}))}
\sin \vartheta (\boldsymbol{x}),  \label{isospin1}
\end{align}
with all the others being zero, 
where $\sigma (\boldsymbol{x})$ and $\vartheta (\boldsymbol{x})$ are the canonical set of the NG mode. 
The ground-state expectation values of these fields must be $\langle \sigma (\boldsymbol{x})\rangle =\sigma _{0}$ 
and $\langle \vartheta (\boldsymbol{x})\rangle=\vartheta_0=-\omega $. 
Since $\rho _{\Phi }\sigma (\boldsymbol{x})$ represents the density while $\vartheta (\boldsymbol{y})$
the angle variable, the following canonical commutation relation holds, 
\begin{equation}
\rho _{\Phi }\left[ \sigma (\boldsymbol{x}),\vartheta (\boldsymbol{y})\right]
=i\delta (\boldsymbol{x}-\boldsymbol{y}).  \label{sigmathetaccr}
\end{equation}
We refer the detailed derivation of the SU(4) isospins (\ref{isospin1}) 
and the canonical commutation relation (\ref{sigmathetaccr}) to Ref.\cite{yhamaetal}.

\section{Effective Hamiltonian for the NG Mode in the CAF phase}\label{appendixB}

We next derive the effective Hamiltonian density and the dispersion relation.
Apart from irrelevant constants, the basic Hamiltonian density for the ground state and the associated NG modes is given by \cite{Ezawa:2013ae}
\begin{align} 
{\mathcal{H}}^{\text{eff}}& =J_{s}^{d}\left( \sum (\partial _{k}
\mathcal{S}_{a})^{2}+(\partial _{k}\mathcal{P}_{a})^{2}
+(\partial _{k}\mathcal{R}_{ab})^{2}\right)  
 +2J_{s}^{-}\left( \sum (\partial _{k}\mathcal{S}_{a})^{2}+(\partial _{k}
\mathcal{P}_{z})^{2}+(\partial _{k}\mathcal{R}_{az})^{2}\right)   \notag \\
& +\rho _{\phi }\left[\epsilon _{\text{cap}}(\mathcal{P}_{z})^{2}
-2\epsilon_{X}^{-}
\left( \sum_a (\mathcal{S}_{a})^{2}
+(\mathcal{R}_{az})^{2}\right) 
 -(-\Delta _{\text{Z}}\mathcal{S}_{z}+\Delta _{\text{SAS}}\mathcal{P}_{x}
\Delta _{\text{bias}}\mathcal{P}_{z})\right],
\label{su4effectivehamiltonian1}
\end{align} 
where $k=x,y$,
$\Delta _{\text{bias}}$ the bias parameter, and
\begin{align}
J_{s}&=J_{s}^{+}+J_{s}^{-}=\frac{1}{16\sqrt{2\pi}}E^0_\text{C},  \quad J_{s}^{d}=J_{s}^{+}-J_{s}^{-} \quad
J^d_s=J_s \left[
-\sqrt{\frac{2}{\pi}}\frac{d}{\ell _{B}}+
\left(
1+\frac{d^2}{\ell _{B}^2}\right)
e^{d^2/2\ell _{B}^2}\text{erfc}\left(d/\sqrt{2}\ell _{B}  \right)
\right], \notag\\
\epsilon _{X}&=\frac{1}{2}\sqrt{\frac{\pi}{2}}E^0_\text{C}, \quad
\epsilon_X^\pm=\frac{1}{2}\left[ 
1\pm  e^{d^2/2\ell _{B}^2}\text{erfc}\left(d/\sqrt{2}\ell _{B} \right)
\right]\epsilon_X, \quad
\epsilon _{D}^{-}=\frac{d}{4\ell _{B}}E^0_\text{C}, \quad
\epsilon _{\text{cap}}=4\epsilon _{D}^{-}-2\epsilon _{X}^{-},\label{spinstifnessdefinition1}
\end{align} 
with $E^0_\text{C}=e^2/4\pi l_B^2$ and $d$ the layer separation.

We introduce the fluctuation fields $\delta \sigma (\boldsymbol{x})$ and $\delta \vartheta (\boldsymbol{x})$ around the ground state
by
\begin{equation}
\sigma (\boldsymbol{x})=\sigma _{0}+\delta \sigma (\boldsymbol{x}),\quad
\vartheta (\boldsymbol{x})=\vartheta_0+\delta \vartheta (\boldsymbol{x}).
\label{NGmode4}
\end{equation}
Note that   $\vartheta_0$ is nothing but the the orientation angle of the in-plane spin component $\omega$ $(\times-1)$
introduced in the main text: See (\ref{inplanespinconfiguration}).
Substituting Eq. \eqref{isospin1} together with (\ref{NGmode4}) into the Hamiltonian density \eqref{su4effectivehamiltonian1}, we obtain 
\begin{equation}
\mathcal{H}_{\text{eff}}=\frac{J_{\vartheta }}{2}\left( \nabla {\delta
\vartheta }\right) ^{2}+\frac{2J_{\sigma }}{\rho _{0}^{2}}\left( \nabla 
\check{\sigma}\right) ^{2}+\frac{2\epsilon _{\text{cap}}^{\nu =1}}{\rho _{0}}\check{\sigma}^{2},  \label{nghamiltonian1}
\end{equation}
where $\epsilon_{\text{cap}}^{\nu =1}=\epsilon_{\text{cap}}-2\epsilon^-_{\text{X}}$, 
$\check{\sigma}(\boldsymbol{x})=\rho_\Phi\delta \sigma(\boldsymbol{x})$ and
\begin{align}
J_{\sigma}=4J_{s}+\frac{(2\sigma_{0}-1)^{2}}{\sigma_{0}(1-\sigma_{0})}J_{s}^{d},\quad
J_{\vartheta}=4{J_{s}^{d}}\sigma_{0}(1-\sigma_{0}).\label{ngstifness}
\end{align}
The Hamiltonian density \eqref{nghamiltonian1} is written in the second-quantized form when the canonical commutation relation 
\eqref{sigmathetaccr} is imposed.
We introduce the annihilation and creation operators,
\begin{align}
&r_{\boldsymbol{k}}=\frac{1}{\sqrt{2}}\left(
\sqrt{G_{\boldsymbol{k}}}\check{\sigma_{\boldsymbol{k}}}+i\frac{1}{\sqrt{G_{\boldsymbol{k}}}}
{\delta\vartheta_{\boldsymbol{k}}}
\right), \qquad
r^\dagger_{\boldsymbol{k}}=\frac{1}{\sqrt{2}}\left(
\sqrt{G_{\boldsymbol{k}}}\check{\sigma}_{\boldsymbol{k}}^\dagger-i\frac{1}{\sqrt{G_{\boldsymbol{k}}}}
{\delta\vartheta^\dagger_{\boldsymbol{k}}}
\right),\label{ngancroperator}
\end{align}
where $\check{\sigma}_{\boldsymbol{k}}$ and 
$\delta{\vartheta}_{\boldsymbol{k}}$ denoting the 
Fourier transforms of the fields  $\check{\sigma}(\boldsymbol{x})$ and $\delta\vartheta(\boldsymbol{x})$,
respectively, and
\begin{align}
G_{\boldsymbol{k}}=\left(\frac{\lambda_\sigma}{\lambda_\vartheta}
\right)^{1/2}, \quad
\lambda_\sigma=\frac{2J_{\sigma}}{\rho^2_{0}}\boldsymbol{k}^{2}
+\frac{2\epsilon_{\text{cap}}^{\nu =1}}{\rho_0}, \quad \lambda_\vartheta=\frac{J_{\vartheta}}{2}\boldsymbol{k}^{2}.
\label{gdefinition}
\end{align}
From \eqref{sigmathetaccr}, we obtain the commutaion relations
\begin{align}
\left[ r_{\boldsymbol{k}}, r^\dagger_{\boldsymbol{k}^\prime}
\right]=\delta(\boldsymbol{k}-\boldsymbol{k}^\prime),\qquad 
\left[\check{\sigma_{\boldsymbol{k}}},
{\delta\vartheta^\dagger_{\boldsymbol{k}^\prime}}
\right]=i\delta(\boldsymbol{k}-\boldsymbol{k}^\prime). \label{ngccrrs}
\end{align}
By using \eqref{ngancroperator}, 
 the Hamiltonian density \eqref{nghamiltonian1} in the momentum space is diagonalized as
\begin{align}
{H}_{\text{R}}&=\int d^2 k E_{\boldsymbol{k}}r^\dagger_{\boldsymbol{k}}r_{\boldsymbol{k}},
\label{EffecHamil2}
\end{align}
where $E_{\boldsymbol{k}}$ is given by
\begin{align}
E_{\boldsymbol{k}}=|\boldsymbol{k}|\sqrt{\frac{2J_{\vartheta}}{\rho_{0}}
\left( \frac{2J_{\sigma}}{\rho_{0}}\boldsymbol{k}^{2}
+2\epsilon_{\text{cap}}^{\nu =1}\right)}
\simeq 2|\boldsymbol{k}|\sqrt{\frac{J_{\vartheta}\epsilon_{\text{cap}}^{\nu =1}}{\rho_{0}}}=\gamma|\boldsymbol{k}|.
\label{nglineardispersion1}
\end{align}
Hence, the NG mode  has the linear dispersion relation.

The  wave length of the NG mode at $\omega_{\text{s}} \sim10$ rad$\cdot$MHz/T is estimated as
\begin{align}
\lambda_{\text{s}}=\frac{2\pi}{k_{\text{s}}}=
2.90454\times 10^{7} \ \text{\AA}, \label{ngmodewavelength}
\end{align}
where $k_{\text{s}}=\hbar\omega_{\text{s}} \gamma^{-1},$
and we have set $l_B=d=230.967$ \AA, or equivalently, 
$\rho_0=0.59669\times10^{-5}$ \AA$^{-2}$, while $\sigma_0=0.3676$, as typical values for QH samples, 
and $k_\text{B}=1.3807\times10^{-23}$ J/K. 
The wavelength \eqref{ngmodewavelength} is about the same size as the sample size, L $\sim$ 100$\mu$m.
Thus the long-wavelength approximation is valid in this case. 

Since  $\check{\sigma}(\boldsymbol{x})$ and $\delta\vartheta(\boldsymbol{x})$ are real fields and  $G_{\boldsymbol{k}}$ are even function of $\boldsymbol{k},$
we obtain the relations $G_{\boldsymbol{k}}=G_{-\boldsymbol{k}}$, $\check{\sigma}^\dagger_{\boldsymbol{k}}=\check{\sigma}_{-\boldsymbol{k}}$ and
$\delta\vartheta^\dagger_{\boldsymbol{k}}=\delta\vartheta_{-\boldsymbol{k}}.$
 Thus, by using them 
the phase field $\delta\vartheta(\boldsymbol{x})$ is described in terms of \eqref{ngancroperator} as
\begin{align}
\delta \vartheta(\boldsymbol{x})&=\int \frac{d^2k}{2\pi}e^{i\boldsymbol{k}\boldsymbol{x}}\left(
-i\sqrt{\frac{G_{\boldsymbol{k}}}{2}}(r_{\boldsymbol{k}}-r^\dagger_{-\boldsymbol{k}})
\right)=\int \frac{d^2k}{2\pi}\left(\kappa_{\boldsymbol{x},\boldsymbol{k}}r_{\boldsymbol{k}}
+\kappa^\ast_{\boldsymbol{x},\boldsymbol{k}}r^\dagger_{\boldsymbol{k}})
\right),\notag\\
\kappa_{\boldsymbol{x},\boldsymbol{k}}&=\sqrt{\frac{G_{\boldsymbol{k}}}{2}}e^{i(\boldsymbol{k}\boldsymbol{x}-\frac{\pi}{2})}, \quad 
\kappa^\ast_{\boldsymbol{x},\boldsymbol{k}}=\sqrt{\frac{G_{\boldsymbol{k}}}{2}}e^{-i(\boldsymbol{k}\boldsymbol{x}-\frac{\pi}{2})}.
\label{kappadefinition}
\end{align}
In the long wave-length limit, we have $e^{i\boldsymbol{k}\boldsymbol{x}}\rightarrow1$, and there is no $\boldsymbol{x}$
dependence in $\kappa_{\boldsymbol{x},\boldsymbol{k}}$. In this limit we just write  it as  $\kappa_{\boldsymbol{k}}$.

\section{Dicke Model in the CAF phase}\label{appendixC}

We proceed to derive the interaction Hamiltonian between the nuclear spins and the NG mode in the CAF phase 
based on \eqref{hyperfine3}. 
We assume that only the nuclear spins  in  one of the layers are  dynamically polarized. 
This is indeed the case in the experiment \cite{Fauziprb}.
Thus we consider the interaction between nuclear spins in the front 
layer and the NG mode $\delta\vartheta$. From now on, we omit the pseudospin index  in the spin density.
First from \eqref{isospin1}, we see that $\mathcal{S}_z$ is dynamically frozen,
because the imbalanced field $\sigma$ has a gap much larger than the thermal energy, and therefore, its excitation is suppressed. 
Thus  the interaction between the nuclear spins and the NG mode is solely expressed by the spin-spin interaction for the in-plane component.
 For  simplicity,  we start from Eq. \eqref{cafspincomponent1} where any limits are not taken. 
 By setting $-\omega\rightarrow\vartheta(\boldsymbol{X}_i)=\vartheta_0+\delta\vartheta(\boldsymbol{X}_i)$, 
and expanding $\mathcal{S}_{a}(\boldsymbol{X}_i)$ in terms of $\delta\vartheta$ up to the linear order, we have
\begin{align}
\mathcal{S}_{x}(\boldsymbol{X}_i)=
S\left(\cos\vartheta_0-\sin\vartheta_0\cdot\delta\vartheta(\boldsymbol{X}_i)
\right)+\mathcal{O}(\delta\vartheta^2),   \quad
\mathcal{S}_{y}(\boldsymbol{X}_i)=
S\left(\sin\vartheta_0+\cos\vartheta_0\cdot\delta\vartheta(\boldsymbol{X}_i)
\right)+\mathcal{O}(\delta\vartheta^2),
\label{cafspincomponent2}
\end{align}
where 
$S=\Delta _{\text{SAS}}\alpha \sqrt{1-\alpha ^{2}}\sqrt{1-\beta ^{2}}/2{\Delta _{0}}$ with
$0\leq S<1$. 
Then, from \eqref{hyperfine3} the hyperfine interaction for the in-plane component becomes
\begin{align}
H_{\text{HF}}&=\sum_{i=1}^N g(\cos\vartheta_0 I_i^x+\sin\vartheta_0 I_i^y)
+\sum_{i=1}^N g\delta\vartheta(\boldsymbol{X}_i)(-\sin\vartheta_0 I_i^x+\cos\vartheta_0 I_i^y),
\label{apphyperfine4}
\end{align}
where $g=\tilde{g}S$.
The first term in \eqref{apphyperfine4} represents the in-plane Knight shift term and 
is much smaller compared with the Larmor frequency $\omega_{\text{s}}$ (see the discussion in Sec. \ref{sec2:hyperfine}). With the same reason, we can neglect the interaction term between the $z$ component 
of nuclear spins and that of electron spins.
Hence we only retain the second term, representing the interaction between the nuclear spins and the NG mode. By introducing $I^{\pm}=I^x\pm i I^y$
and using \eqref{kappadefinition}, we have
\begin{align}
H_{\text{HF}}&=\frac{g}{2}\sum_{i=1}^N (\tilde{I}_i^++\tilde{I}_i^-)
\int \frac{d^2k}{2\pi}\left(\kappa_{\boldsymbol{X}_i,\boldsymbol{k}}r_{\boldsymbol{k}}
+\kappa^\ast_{\boldsymbol{X}_i,\boldsymbol{k}}r^\dagger_{\boldsymbol{k}})
\right),
\label{supphyperfine5}
\end{align}
where  $\tilde{I}^{\pm}=e^{\mp i(\vartheta_0+\pi/2)}{I}^{\pm}$ is the rotated in-plane nuclear spin. 
We just write them as  ${I}^{\pm}$ in the rest of this Appendix. By using the rotating-wave  approximation,
we obtain
\begin{align}
H_{\text{SR}}=\frac{g}{2}\sum_{i=1}^N (I^+_i R_i+ I^-_i R^\dagger_i), \qquad
R_i=\int \frac{d^2k}{2\pi}\kappa_{\boldsymbol{X}_i,\boldsymbol{k}}r_{\boldsymbol{k}}, \quad
R_i^\dagger=\int \frac{d^2k}{2\pi}\kappa^\ast_{\boldsymbol{X}_i,\boldsymbol{k}}r^\dagger_{\boldsymbol{k}}.
\label{supphyperfine6}
\end{align}
In the long-wavelength limit the $i$ dependence disappears from $R_i$ and $R_i^\dagger.$
Thus we obtain the interaction Hamiltonian for the nuclear spins and the NG mode in the CAF phase,
\begin{align}
H_{\text{SR}}&=\frac{g}{2}\sum_{i=1}^N (I^+_i R+ I^-_i R^\dagger)
=\frac{g}{2} (J^+ R+ J^- R^\dagger).
\label{supphyperfine7}
\end{align}
By combining the Larmor-precision Hamiltonian \eqref{Larmorprecision}, 
the effective Hamiltonian for the NG mode \eqref{EffecHamil2}, and the interaction Hamiltonian \eqref{supphyperfine7}, 
we have the Dicke model in the CAF phase.

\end{document}